%% file: Main-part.tex
\documentclass[conference, letterpaper]{IEEEtran}
\IEEEoverridecommandlockouts
\usepackage{cite}
\usepackage{amsmath,amssymb,amsfonts}
\usepackage{algorithm,algorithmic}
\usepackage{graphicx}
\usepackage{textcomp}
\usepackage{amsmath}  
\usepackage{amsfonts}  
\usepackage{amssymb}  
\usepackage{soul}
\usepackage{enumitem}
\usepackage{xcolor}
\usepackage{booktabs}
\usepackage{cuted}
\usepackage{svg}
\usepackage{tcolorbox}
\usepackage{bm}
\usepackage[letterpaper, right=0.626in, left=0.624in, top=0.73in, bottom=1.09in]{geometry}

\def\BibTeX{{\rm B\kern-.05em{\sc i\kern-.025em b}\kern-.08em
    T\kern-.1667em\lower.7ex\hbox{E}\kern-.125emX}}

\DeclareRobustCommand*{\IEEEauthorrefmarknum}[1]{%
    \raisebox{0pt}[0pt][0pt]{\textsuperscript{\footnotesize\ensuremath{#1}}}}

\makeatletter
\def\@fnsymbol#1{\ensuremath{\ifcase#1\or *\or \dagger\or \ddagger\or \mathsection\or \mathparagraph\or \|\or **\or \dagger\dagger\or \ddagger\ddagger \fi}}
\makeatother

\begin{document}

\title{Joint Antenna Position and Beamforming Optimization with Self-Interference Mitigation in Movable Antenna Aided ISAC System}
\author{

\IEEEauthorblockN{Size Peng\IEEEauthorrefmarknum{1}\IEEEauthorrefmark{1}, Cixiao Zhang\IEEEauthorrefmarknum{1}\IEEEauthorrefmark{1}, Yin Xu\IEEEauthorrefmarknum{1}, Qingqing Wu\IEEEauthorrefmarknum{2}, Lipeng Zhu\IEEEauthorrefmarknum{3}, Xiaowu Ou\IEEEauthorrefmarknum{1} and Dazhi He\IEEEauthorrefmarknum{1}}
\IEEEauthorblockA{
\IEEEauthorrefmarknum{1}Cooperative Medianet Innovation Center (CMIC), Shanghai Jiao Tong University, Shanghai 200240, China.\\
\IEEEauthorrefmarknum{2}Department of Electronic Engineering, Shanghai Jiao Tong University, Shanghai, China.\\
\IEEEauthorrefmarknum{3}Department of Electrical and Computer Engineering, National University of Singapore, Singapore 117583.\\
Email: sjtu2019psz@sjtu.edu.cn, cixiaozhang@sjtu.edu.cn, xuyin@sjtu.edu.cn, qingqingwu@sjtu.edu.cn,\\ zhulp@nus.edu.sg, xiaowu\_ou@sjtu.edu.cn, hedazhi@sjtu.edu.cn
}

    \thanks{   
    This paper is supported in part by National Key R\&D Project of China (2023YFB2906201), National Natural Science Foundation of China Program(62422111, 62271316, 62371291), the Fundamental Research Funds for the Central Universities and Shanghai Key Laboratory of Digital Media Processing (STCSM 18DZ2270700).
    
    \IEEEauthorrefmark{1}These authors contributed equally to this work.
    
    The corresponding author is Yin Xu (e-mail: xuyin@sjtu.edu.cn).}
}

\maketitle

\begin{abstract}
Movable antennas (MAs) have shown significant potential in improving the performance of integrated sensing and communication (ISAC) systems. However, their application in integrated and cost-effective full-duplex (FD) monostatic systems remains underexplored. To bridge this research gap, we develop an MA-ISAC model within an FD monostatic framework, where the self-interference channel is modeled as a function of the antenna position vectors under the near-field channel condition. This model enables antenna position optimization for maximizing the weighted sum of communication capacity and sensing mutual information. The resulting optimization problem is non-convex making it challenging to solve optimally. To address this, we employ the fractional programming (FP) method and propose an alternating optimization (AO) algorithm that jointly optimizes the beamforming and antenna positions at the transceivers. Specifically, closed-form solutions for the transmit and receive beamforming matrices are derived using the Karush–Kuhn–Tucker (KKT) conditions, and a novel coarse-to-fine grained searching (CFGS) approach is used to determine high-quality sub-optimal antenna positions. Numerical results demonstrate that with strong self-interference cancellation (SIC) capabilities, MAs significantly enhance the overall performance and reliability of the ISAC system when utilizing our proposed algorithm, compared to conventional fixed-position antenna designs.
\end{abstract}

\input{Chapter/Introduction}
\input{Chapter/system-model}

\input{Chapter/proposed-solution}



\bibliographystyle{IEEEbib}
\bibliography{IEEEabrv,refs}

\end{document}

%% file: Chapter/Introduction.tex
\section{Introduction}
The increasing demand for reliable sensing and efficient communication has sparked significant interest in Integrated Sensing and Communication (ISAC) technologies. ISAC aims to merge communication and sensing functions within a single system, utilizing the same frequency bands and hardware resources. This integration improves spectral resource utilization, reduces hardware costs, and simplifies system complexity, positioning ISAC as a highly promising and efficient approach for modern wireless networks. Recent studies \cite{ISACbeamforming,survey-isac2} show that ISAC systems can achieve notable improvements in spectral efficiency compared to traditional systems that separate communication and sensing functions. As wireless networks advance toward 6G and beyond, ISAC is expected to be crucial in meeting the growing demands for extremely higher data rates, lower latency, and improved connectivity.

Beamforming design is critical in both multiple-input multiple-output (MIMO) communication and sensing systems due to its capability of array signal processing. However, traditional systems with fixed and equally spaced antennas cannot fully exploit the spatial degrees of freedom (DoF) offered by multiple antennas. To address this limitation, a movable antenna (MA) system, also known as a fluid antenna system, has been proposed. This system can flexibly adjust antenna positions and thus capture the spatial variations of wireless channels for improving communication/sensing performance \cite{R1}. The superiority of MA systems over conventional fixed-position antenna (FPA) systems has been widely studied and validated in terms of flexible beamforming \cite{lyu2024flexible}, spatial multiplexing \cite{zhumulti}, index modulation \cite{guo2024fluid}, and other aspects. 


In the context of ISAC, the full-duplex (FD) monostatic setup is practically appealing for automotive and Internet of Things (IoT) applications due to its seamless integration, cost-effectiveness, and efficient use of the spectrum \cite{ISAC-survey}. Although some existing studies have explored the use of MAs in the ISAC scenario \cite{lyu2024flexible,FluidISAC,hao2024fluid,ma2024movableantennaaidedsecuretransmission}, an FD monostatic scenario remains under-researched. This paper pioneers the investigation of the effectiveness of MAs in this important yet challenging setup. In particular, self-interference cancellation (SIC) is a critical issue under this setup. High performance can only be achieved with strong SIC capability. Unlike physical isolation methods, active suppression of self-interference (SI) can be achieved through Tx and Rx beamforming \cite{SIC_survey,shi_robust_2022}. Motivated by \cite{dai_2023,RISFD,shi_robust_2022}, we model the SI channel as a function of the positions of the transmit and receive antennas under the near-field channel condition, allowing for more precise control and reduction of SI by antenna position optimization.

To characterize the trade-off between communication and sensing, we maximize the weighted sum of communication rate and sensing mutual information (MI). The beamforming matrix and antenna positions at Tx and Rx are optimized using the alternating optimization (AO) method. Specifically, we propose a coarse-to-fine grained searching (CFGS) algorithm to determine a high-quality sub-optimal antenna positions. Our contributions are briefly summarized as follows
\begin{itemize}
    \item We model the FD monostatic MA-ISAC system, including the SI channel characterized by antenna position vectors at the transmitter and receiver. This enables the strategic positioning of the MA to mitigate interference.
    \item We propose an efficient algorithm to jointly optimize the transmit and receive beamforming matrices along with the Tx/Rx antenna positions.
    \item Numerical results demonstrate the effectiveness of our algorithm in enhancing performance of the considered FD monostatic ISAC system.
\end{itemize}

\textit{Notations:} $\mathbf{x}(n)$, $\mathbf{x}^{T}$, $\mathbf{x}^{*}$, $\text{Tr}(\mathbf{X})$, $(\mathbf{X})^{-1}$ and $[\mathbf{X}]_{j,i}$ denote the $n^{th}$ entry of $\mathbf{x}$, the transpose of $\mathbf{x}$, the conjugate of $\mathbf{x}$, the trace of $\mathbf{X}$, the inverse of $\mathbf{X}$ and the entry in the $j^{th}$ row and $i^{th}$ column of the matrix $\mathbf{X}$, respectively. 

%% file: Chapter/system-model.tex
\section{System Model}
\begin{figure}[t]
    \centering
    \includegraphics[width=0.35\textwidth]{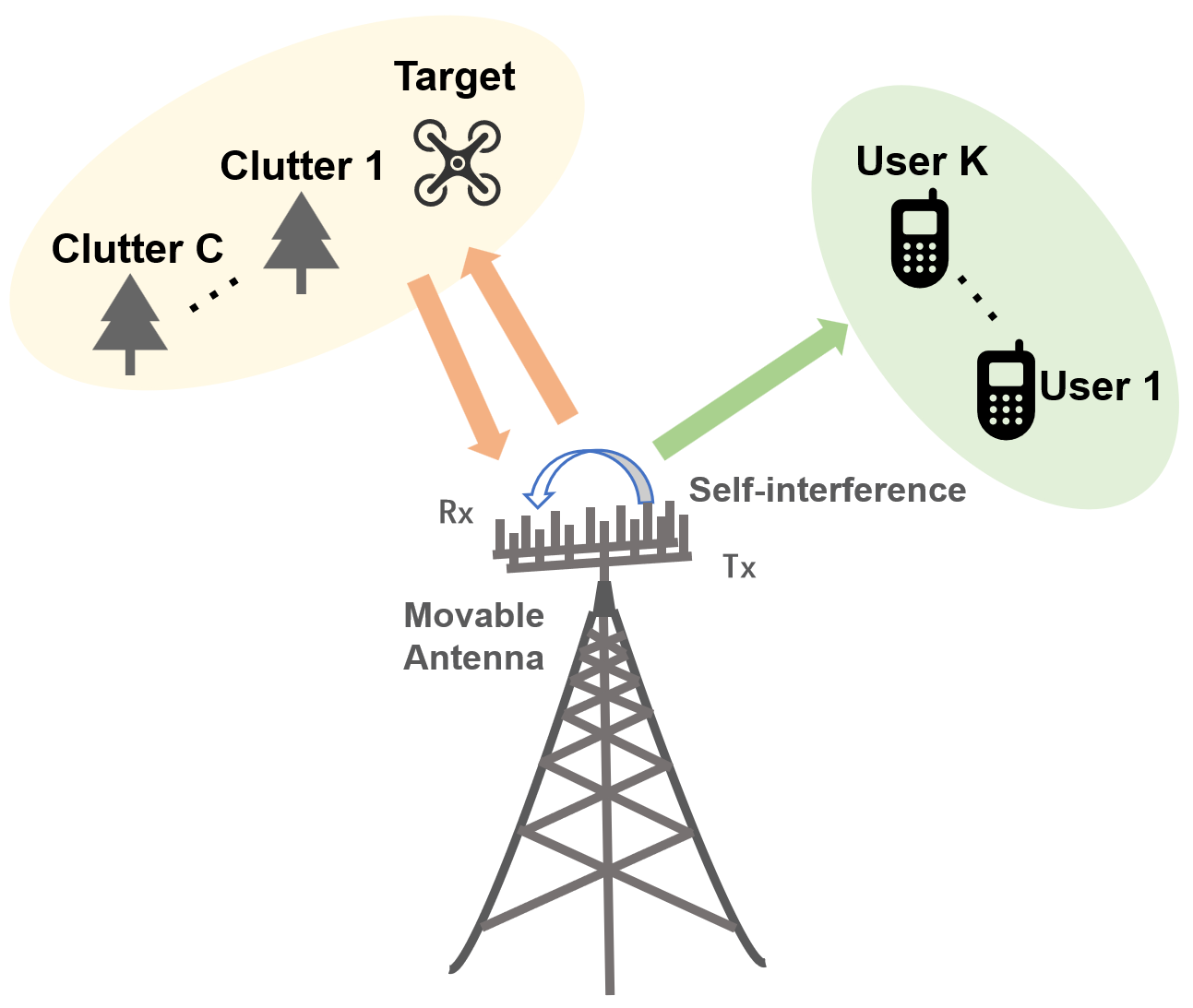}
    \caption{System model of the monostatic MA-ISAC system.}
    \vspace{-0.5cm}
    \label{fig:system model}
\end{figure}

In this paper, we consider an FD monostatic base station (BS) with $N_T$ linear Tx-MAs and $N_R$ linear Rx-MAs surrounded by $K$ users, $C$ clutters and a sensing target, as illustrated in Fig. \ref{fig:system model}. The antenna positions in Tx and Rx can be adjusted flexibly within two parallel line segments $[X_{\min},X_{\max}]$ and  $[Y_{\min},Y_{\max}]$, respectively. 
\subsection{Channel Model}
The antenna position vectors of the Tx-MA array and the Rx-MA array are denoted as $\mathbf{x}=[x_1,..., x_{N_T}]^T$ and $\mathbf{y}=[y_1,..., y_{N_R}]^T$. 
According to the far-field response model in \cite{10318061}, the transmit steering vector of the Tx-MA array is
\begin{equation}
\scalebox{0.86}{$
\mathbf{a}_{k,l}(\mathbf{x}) = \begin{bmatrix} e^{j\frac{2\pi }{\lambda} x_1\cos(\theta_{k,l})},\cdots, e^{j\frac{2\pi }{\lambda} x_{N_T}\cos(\theta_{k,l})} \end{bmatrix}^T\in \mathbb{C}^{N_T},$
}
\label{steering_vector}
\end{equation}
where $\lambda$ is the carrier wavelength and $\theta_{k,l}$ denotes the angle of departure of the $l^{th}$ path for the $k^{th}$ user. Denote $\eta_k$ as the free-space path loss and $\rho_{k,l}$ as the channel gain coefficient experienced by the $l^{th}$ path for the $k^{th}$ user. $L_p$ is the number of rays. Therefore, the channel between the BS and the $k^{th}$ user is
\begin{equation}
    \mathbf{h}_k(\mathbf{x}) = \sqrt{\frac{\eta_k}{L_p}} \sum\limits_{l=1}^{L_p} \rho_{k,l} \mathbf{a}_{k,l}(\mathbf{x})\in \mathbb{C}^{N_T}.
\end{equation}
Additionally, the receive steering vector of the Rx-MA array in the direction of $\psi$ is given by
\begin{equation}
\scalebox{0.95}{$
    \mathbf{b}(\mathbf{y}) = \begin{bmatrix} e^{j\frac{2\pi }{\lambda} y_1\cos(\psi)},\cdots,e^{j\frac{2\pi }{\lambda} y_{N_R}\cos(\psi)} \end{bmatrix}^T\in \mathbb{C}^{N_R}.
    $}
\end{equation}
Therefore, channels for sensing target and $c^{th}$ clutter are
\begin{equation}
    \mathbf{h}_s(\mathbf{x,y})=\sqrt{\eta_s} \alpha_{s} \mathbf{a}_{s}(\mathbf{x})\mathbf{b}_{s}^H(\mathbf{y})\in \mathbb{C}^{N_T\times N_R},
\end{equation}
\begin{equation}
    \mathbf{h}_c(\mathbf{x,y})=\sqrt{\eta_c} \alpha_{c} \mathbf{a}_{c}(\mathbf{x})\mathbf{b}_{c}^H(\mathbf{y})\in \mathbb{C}^{N_T\times N_R},
\end{equation}
where the complex coefficients $\alpha_{s}$ and $\alpha_{c}$ represent the radar cross section (RCS) of the sensing target and the $c^{th}$ clutter, respectively. Likewise, $\eta_s$ and $\eta_c$ denote the free-space path losses along the paths of the target and the $c^{th}$ clutter, respectively. The transmit steering vectors corresponding to the directions of the target and the $c^{th}$ clutter are represented by $\mathbf{a}_s$ and $\mathbf{a}_c$, respectively. $\mathbf{b}_s$ and $\mathbf{b}_c$ are the corresponding receive steering vectors.

\begin{figure}[t]
    \centering
    \includegraphics[width=0.75\linewidth]{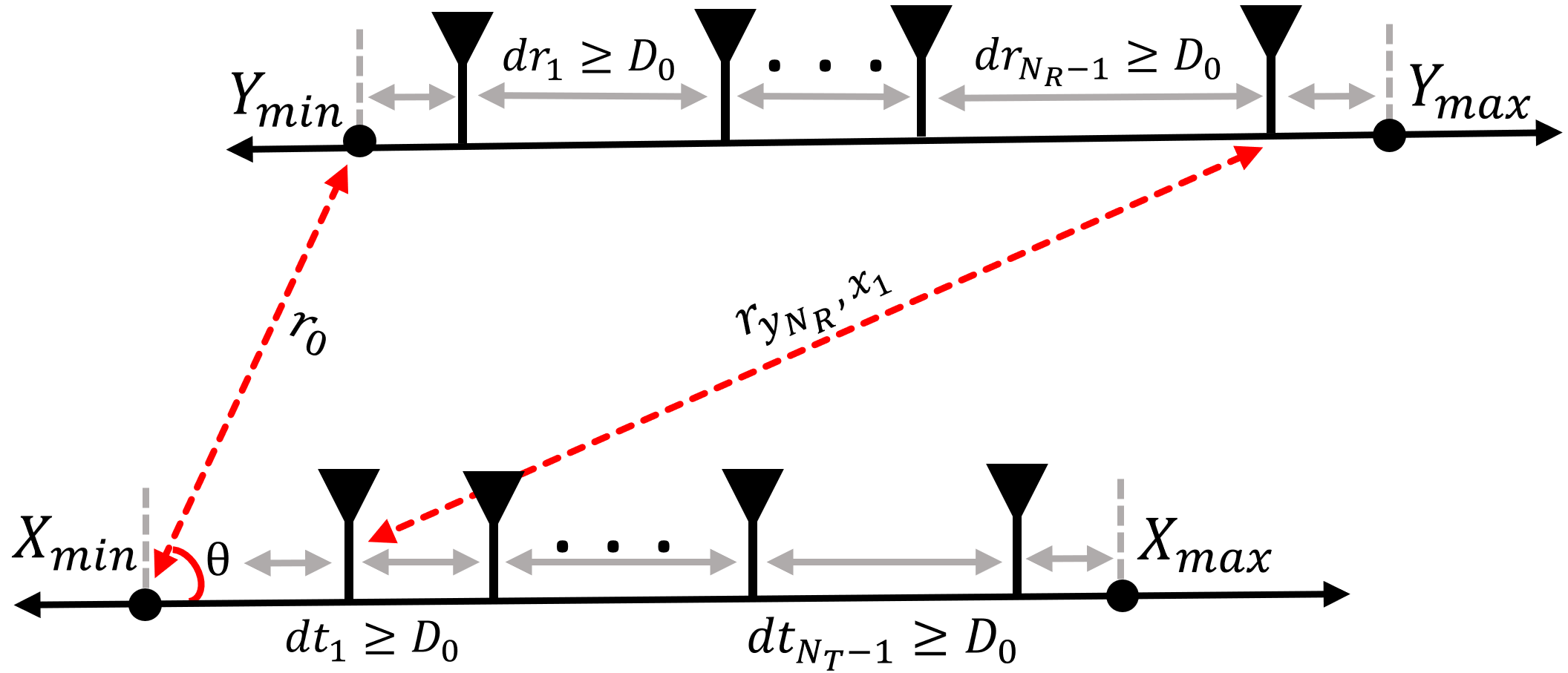}
    \caption{Movable antenna model at Tx and Rx.}
    \vspace{-0.5cm}
    \label{fig:antenna spacing}
\end{figure}

With increasing apertures for both transmit and receive arrays and higher operating frequency (resulting in a reduction in the wavelength $\lambda$), the distance between the transmitter and receiver becomes less than the MIMO Rayleigh distance of $\frac{(D_T + D_R)^2}{\lambda}$\cite{dai_2023}. Consequently, different from the far-field channel, the SI channel should be modeled as a near-field channel \cite{RISFD,shi_robust_2022}. We define $r_{y_j,x_i}$ as the distance between the $j^{th}$ receive antenna and the $i^{th}$ transmit antenna. $r_0$ represents the distance between $Y_{\min}$ and $X_{\min}$, as demonstrated in Fig. \ref{fig:antenna spacing}.
The distance can be expressed as
\begin{equation}
\scalebox{0.83}{$
    r_{y_j,x_i} = \sqrt{
        r_0^2 + x_i^2 + y_j^2 + 2r_0y_j\cos(\theta) - 
        2r_0x_i\cos\theta - 2y_jx_i  
    }.
$}
\end{equation}
Let $\boldsymbol{\eta}_{\text{SI}}$ denote the free-space path loss matrix and $G_l$ antenna gain. The expression of the entry in the $j^{th}$ row and  the $i^{th}$ column of SI channel matrix $\mathbf{H}_{\text{SI}}\in \mathbb{C}^{N_R\times N_T}$ is given by
\begin{equation}
[\mathbf{H}_{\text{SI}}]_{j,i}=\left(\sqrt{[\boldsymbol{\eta}_{\text{SI}}]_{j,i}}\ e^{-j \frac{2\pi}{\lambda} r_{ y_j,x_i}}\right),
\end{equation}
 where $
    [\boldsymbol{\eta}_{\text{SI}}]_{j,i}=\frac{G_l}{4}\bigg[\big(\frac{\lambda}{2\pi r_{y_j,x_i}}\big)^2-\big(\frac{\lambda}{2\pi r_{y_j,x_i}}\big)^4+\big(\frac{\lambda}{2\pi r_{y_j,x_i}}\big)^6\bigg].
     $




\subsection{Signal Model}
Let $\mathbf{s}=[s_1,s_2,...,s_K]^T$, $\mathbb{E}\{ \mathbf{s} \mathbf{s}^H \} = \mathbf{I}$, denote the signals for $K$ users, which is used for both communication and sensing. The transmit and receive beamforming matrices are
\begin{equation}
    \mathbf{F}=[\mathbf{f}_1,\mathbf{f}_2,...,\mathbf{f}_K]\in \mathbb{C}^{N_T \times K},
\end{equation}
\vspace{-0.5cm}
\begin{equation}
    \mathbf{w}=[w_1,w_2,...,w_{N_R}]^H\in \mathbb{C}^{N_R}.
\end{equation}
Thus, the baseband received signal at the $k^{th}$ user is expressed as
\vspace{-0.15cm}
\begin{equation}
    r_k = \mathbf{h}_k^H(\mathbf{x})\mathbf{f}_ks_k + \mathbf{h}_k^H\sum\limits_{j=1, j\neq k}^{K} \mathbf{f}_j s_j + n_k. 
\end{equation}
The signal-to-interference-plus-noise ratio (SINR) at the $k^{th}$ user can be derived as
\begin{equation}
\text{SINR}_k = \frac{|\mathbf{h}_k^H(\mathbf{x})\mathbf{f}_k|^2}{\sum\nolimits_{j=1, j \neq k}^{K} |\mathbf{h}_k^H(\mathbf{x})\mathbf{f}_j|^2 + \sigma_k^2},
\end{equation}
where receive noise for $k^{th}$ user $n_k \sim \mathcal{C}\mathcal{N}(0,\sigma_k^2)$. Consequently, the communication rate for the $k^{th}$ user is
\begin{equation}
R_k=\log(1+\text{SINR}_k).\label{Communication Rate}
\end{equation}
The received signal at the BS with receive beamforming can be expressed as
\begin{equation}
\scalebox{0.9}{$
\begin{aligned}
        r_s =&  \sqrt{\eta_s}\alpha_s \mathbf{w}^H\mathbf{b}_s(\mathbf{y}) \mathbf{a}_s^H(\mathbf{x}) \mathbf{F} \mathbf{s} 
        +\mathbf{w}^H\mathbf{H_{\text{SI}}}(\mathbf{x}, \mathbf{y}) \mathbf{F} \mathbf{s}\\
        &+ \sum_{c=1}^C \sqrt{\eta_c}\alpha_c\mathbf{w}^H\mathbf{b}_c(\mathbf{y}) \mathbf{a}_c^H(\mathbf{x}) \mathbf{F} \mathbf{s} 
         + \mathbf{w}^H\mathbf{n}_s,
\end{aligned}
$}
\end{equation}
where receive noise $\mathbf{n}_s \sim \mathcal{C}\mathcal{N}(0,\sigma_s^2\mathbf{I}_{N_R})$. Thus, the signal-to-clutter-plus-noise ratio (SCNR) at the BS is  expressed as
\begin{equation}
\scalebox{0.9}{$
\text{SCNR} = \frac{||\sqrt{\eta_s}\alpha_s\mathbf{w}^H \mathbf{b}_s(\mathbf{y})\mathbf{a}_s^H(\mathbf{x}) \mathbf{F}||^2}{\sum\limits_{c=1}^{C} ||\sqrt{\eta_c}\alpha_c \mathbf{w}^H \mathbf{b}_c(\mathbf{y}) \mathbf{a}_c^H(\mathbf{x}) \mathbf{F}||^2+||\mathbf{w}^H \bold{H}_{\text{SI}}(\mathbf{x},\mathbf{y})\mathbf{F}||^2+ ||\mathbf{w}||^2 \sigma_s^2} .
$}
\end{equation}
According to \cite{lyu2024flexible,MI}, the sensing MI can be expressed as
\begin{equation}
R_s = \log(1 + \text{SCNR}).\label{Mutual information}
\end{equation}
\subsection{Problem Formulation}
To balance the communication and sensing performance, we aim to maximize the weighted sum of communication rate and sening MI in (\ref{Communication Rate}) and (\ref{Mutual information}). Hence, the optimization problem is formulated as
\vspace{-0.1cm}
\begin{subequations}
\begin{align}
(\text{P1}) \max_{\mathbf{F}, \mathbf{x}, \mathbf{y}, \mathbf{w}}  \mathcal{G}(\mathbf{F}, \mathbf{x}, \mathbf{y}, \mathbf{w}) = &\varpi_c \sum\nolimits_{k=1}^{K} R_k + \varpi_s R_s \label{eq:originalG} \\
\quad \text{s.t.} \quad  \text{Tr}(\mathbf{F}^H \mathbf{F})& \leq P_0,  \label{eq:F}  \\
\quad X_{\min} \leq x_i \leq X_{\max}, Y_{\min} &\leq y_j \leq  Y_{\max},  \forall i,j \label{eq:xy1} \\
|x_i - x_{\hat{i}}| \geq D_0, |y_j - y_{\hat{j}}|& \geq D_0, i\neq\hat{i}, j\neq\hat{j} \label{eq:xy2},
\end{align}
\end{subequations}
where $P_0$ is the maximum transmit power and $D_0$ is the minimum separation distance between each pair of antennas at Tx or Rx for avoiding the coupling effect. The weighted factors $\varpi_c$ and $\varpi_s$ satisfy $\varpi_c+\varpi_s=1$.

%% file: Chapter/proposed-solution.tex
\begin{figure*}[t]
\vspace{-0.6cm}
    \begin{align}
        &\max_{\mathbf{F}, \mathbf{x}, \mathbf{y}, \mathbf{w}, \boldsymbol{\mu},\boldsymbol{\xi}^c,\boldsymbol{\xi}^s}\hat{\mathcal{G}}(\mathbf{F}, \mathbf{x}, \mathbf{y}, \mathbf{w}, \boldsymbol{\mu},\boldsymbol{\xi}^c,\boldsymbol{\xi}^s) = \varpi_c \sum\limits_{k=1}^{K} \log(1+\mu_k) + \varpi_s \log(1+\mu_{K+1}) -\varpi_c \sum\limits_{k=1}^{K} \mu_k - \varpi_s \mu_{K+1}  \notag\\
        &+ \varpi_c \sum\limits_{k=1}^{K} \Bigg[2\sqrt{1+\mu_k} \text{Re}\{\xi^c_k \mathbf{h}_k^H(\mathbf{x}) \mathbf{f}_k\} - |\xi^c_k|^2 \left(\sum_{j=1}^{K} |\mathbf{h}_k^H(\mathbf{x}) \mathbf{f}_j|^2 + \sigma_k^2\right)\Bigg] +\varpi_s \Bigg[ 2\sqrt{1+\mu_{K+1}} \text{Re}\{\sqrt{\eta_s}\alpha_s \mathbf{w}^H \mathbf{b}_s(\mathbf{y}) \mathbf{a}_s^H(\mathbf{x}) \mathbf{F} \boldsymbol{\xi}^s\}\notag \\
        &- ||\boldsymbol{\xi}^s||^2 \left(\sum_{c=1}^C ||\sqrt{\eta_c}\alpha_c \mathbf{w}^H  \mathbf{b}_c(\mathbf{y}) \mathbf{a}_c^H(\mathbf{x}) \mathbf{F}||^2 + ||\mathbf{w}^H \mathbf{H}_{\text{\text{SI}}}(\mathbf{x},\mathbf{y}) \mathbf{F}||^2 + ||\sqrt{\eta_s}\alpha_s \mathbf{w}^H \mathbf{b}_s(\mathbf{y}) \mathbf{a}_s^H(\mathbf{x}) \mathbf{F}||^2 + ||\mathbf{w}^H||^2 \sigma_s^2\right)\Bigg].
        \label{convexG}
    \end{align}
\hrulefill
\end{figure*}
\section{Proposed Solution}
\vspace{-0.1cm}
It is challenging to solve (\text{P1}) directly because the optimization function (\ref{eq:originalG}) is non-convex w.r.t. $\mathbf{F}$, $\mathbf{w}$, $\mathbf{x}$, and $\mathbf{y}$. To address this problem, we employ the fractional programming (FP) approach \cite{FPmethod}. Auxiliary variables, including $\boldsymbol{\mu}=[\mu_1,\mu_2,...,\mu_{K+1}]^T$, $\boldsymbol{\xi}^c=[{\xi}^c_1,{\xi}^c_2,...,{\xi}^c_K]^T$, and $\boldsymbol{\xi}^s=[{\xi}^s_1,{\xi}^s_2,...,{\xi}^s_K]^T$, are introduced to transform the objective function in (\ref{eq:originalG}) into an equivalent convex form $\hat{\mathcal{G}}$ in (\ref{convexG}). We then present an AO algorithm in which the antenna positions, as well as the transmit and receive beamforming matrices, along with the auxiliary variables, are updated in turn while the other parameters remain fixed until reaching convergence.
\vspace{-0.5cm}

\subsection{Transmit and Receive Beamforming Optimization}
We aim to optimize the transmit and receive beamforming matrix $\mathbf{F}$ and $\mathbf{w}$ with fixed $\mathbf{x}, \mathbf{y}$ and auxiliary variables. Firstly, we formulate the subproblem for $\mathbf{F}$ as follows
\begin{equation}
    (\text{SP.1}) \max_{\mathbf{F}}\hat{\mathcal{G}}(\mathbf{F}|\mathbf{x}, \mathbf{y}, \mathbf{w}, \boldsymbol{\mu},\boldsymbol{\xi}^c,\boldsymbol{\xi}^s)~\quad \\ \text{s.t.} \ \text{(\ref{eq:F})}.
\end{equation}
Since \(\hat{\mathcal{G}}\) is a convex function w.r.t. $\mathbf{F}$, we can employ the Lagrange dual method to obtain the closed-form expression of \(\mathbf{F}\). The Lagrangian function is defined as \(\mathcal{L}(\mathbf{F}, \tau) = -\hat{\mathcal{G}}(\mathbf{F}|\mathbf{x}, \mathbf{y}, \mathbf{w}, \boldsymbol{\mu}, \boldsymbol{\xi}^c, \boldsymbol{\xi}^s) + \tau \left(\text{Tr}(\mathbf{F}^H\mathbf{F}) - P_0\right)\). The Lagrangian dual problem is then characterized by the following Karush–Kuhn–Tucker (KKT) 
 conditions
\begin{subequations}
    \begin{align}
        \frac{\partial \mathcal{L}(\mathbf{F}, \tau)}{\partial \mathbf{F}} &= 0, \label{eq:gradientF}\\
        \text{Tr} \left(\mathbf{F}^H\mathbf{F}\right) - P_0 &\leq 0, \\
        \tau &\geq 0, \\
        \tau \left(\text{Tr} \left(\mathbf{F}^H\mathbf{F}\right) - P_0\right) &= 0,
    \end{align}
\end{subequations}
which yields an optimal solution for $\mathbf{F}$ in closed form, with the $k^{th}$ column given by 
\begin{equation}
\bold{f}_k(\tau) = \bigg(\big( \boldsymbol{\Lambda}_k^{T} + \tau \mathbf{I} \big)^{-1}\bigg)^{*} \boldsymbol{\varphi}_k,
\vspace{-0.2cm}
\label{precoding_update}
\end{equation}
where 
\begin{equation}
\scalebox{0.82}{$
\begin{aligned}
    &\boldsymbol{\Lambda}_k = \varpi_s ||\boldsymbol{\xi}^s||^2 \bigg\{\eta_s|\alpha_s|^2 \left(\mathbf{w}^H \mathbf{b}_s(\mathbf{y}) \mathbf{a}_s^H(\mathbf{x}) \right)^H \left(\mathbf{w}^H \mathbf{b}_s(\mathbf{y}) \mathbf{a}_s^H(\mathbf{x}) \right)\\
    &\quad\quad\quad+ \sum_{c=1}^C \eta_c|\alpha_c|^2 \left(\mathbf{w}^H  \mathbf{b}_c(\mathbf{y}) \mathbf{a}_c^H(\mathbf{x}) \right)^H \left(\mathbf{w}^H  \mathbf{b}_c(\mathbf{y}) \mathbf{a}_c^H(\mathbf{x}) \right)  \\
    &\quad\quad\quad+ \left(\mathbf{w}^H \bold{H}_{\text{SI}}(\mathbf{x},\mathbf{y})\right)^H \left(\mathbf{w}^H \bold{H}_{\text{SI}}(\mathbf{x},\mathbf{y})\right)\bigg\} +\varpi_c|\xi^c_k|^2 \mathbf{h}_k\mathbf{h}_k^H,\\ 
    &\boldsymbol{\varphi}_k=\varpi_c\sqrt{1+\mu_k}\xi_k^{c*}\mathbf{h}_k(\mathbf{x})\nonumber\\ &\quad\quad\quad+\varpi_s\sqrt{1+\mu_{K+1}}\sqrt{\eta_s}\alpha_s^{*}\boldsymbol{\xi}^{s*}_k\mathbf{a}_s(\mathbf{x})\mathbf{b}^H_s(\mathbf{y})\mathbf{w}.
\end{aligned}
$}
\end{equation}

Similarly, we can derive the closed form of $\mathbf{w}$ as follows
\begin{equation}
    \mathbf{w}=\big(\bm{\Psi}^{-1}\big)^*
    \bm{\gamma},
    \label{wr_update}
\end{equation}
where 


\begin{equation}
\scalebox{0.82}{$
\begin{aligned}
    &\bm{\Psi} = ||\boldsymbol{\xi}^s||^2 \bigg\{ \sum\nolimits_{c=1}^{C}\eta_c|\alpha_c|^2\Big( \mathbf{b}_c(\mathbf{y})\mathbf{a}_c^H(\mathbf{x})\mathbf{F}\Big)\Big( \mathbf{b}_c(\mathbf{y})\mathbf{a}_c^H(\mathbf{x})\mathbf{F}\Big)^H \nonumber\\
    &\quad\quad\quad + \eta_s|\alpha_s|^2\Big(\mathbf{b}_s(\mathbf{y})\mathbf{a}_s^H(\mathbf{x})\mathbf{F}\Big)\Big(\mathbf{b}_s(\mathbf{y})\mathbf{a}_s^H(\mathbf{x})\mathbf{F}\Big)^H \nonumber\\
    &\quad\quad \quad+\Big(\mathbf{H}_{\text{SI}}(\mathbf{x},\mathbf{y})\mathbf{F}\Big)\Big(\bold{H}_{\text{SI}}(\mathbf{x},\mathbf{y})\mathbf{F}\Big)^H + \sigma_s^2 \mathbf{I} \bigg\},\\
    &\bm{\gamma}=\sqrt{(1+\mu_{K+1})\eta_s}\alpha_s\mathbf{b}_s(\mathbf{y})\mathbf{a}_s^H(\mathbf{x})\mathbf{F}\boldsymbol{\xi}^s.
\end{aligned}
$}
\end{equation}

To address $\tau$ in the complementary slackness condition, we draw inspiration from \cite{lambda,lyu2024flexible} and employ the bisection method to select the appropriate dual variable, which is detailed in \textbf{Algorithm \ref{algo1}}.

\begin{algorithm}[t]
    \renewcommand{\algorithmicrequire}{\textbf{Initialization:}}
	\renewcommand{\algorithmicensure}{\textbf{Output:}}
    \caption{Iterative optimization for transmit and receive beamforming matrices.}
    \label{algo1}
    \begin{algorithmic}[1]
        \REQUIRE Choose the upper bound and lower bound of $\tau$ as $\tau_{\text{max}}$ and $\tau_{\text{min}}$, tolerence $\epsilon$, power limit $P_{0}$; randomly initial $\boldsymbol{\xi}_{s}$, $\boldsymbol{\xi}_{c}$, $\boldsymbol{\mu}$, $\mathbf{w}$, set iteration index $i=1$.
        \REPEAT
            \REPEAT
                \STATE Compute $\tau=(\tau_{\text{max}}+\tau_{\text{min}})/2$. 
                \STATE Update precoding matrix $\mathbf{F}^{(i)}$ as (\ref{precoding_update}).  
                \STATE Compute power $P$ of precoding matrix $\mathbf{F}^{(i)}$. 
                \STATE \textbf{if} { $P > P_{0}$ } \textbf{then} $\tau_{\text{min}}=\tau$ \textbf{else} $\tau_{\text{max}}=\tau$.
            \UNTIL{$\left| P - P_{0} \right|< \epsilon$}
            \STATE Update $\mathbf{w}^{(i)}, \boldsymbol{\xi}_{c}^{(i)}, \boldsymbol{\xi}_{s}^{(i)}, \boldsymbol{\mu}^{(i)}$  as (\ref{wr_update}), (\ref{ksic_update}), (\ref{ksis_update}), (\ref{mu_dupdate}) seperately. Set iteration index $i=i+1$
        \UNTIL{the value of objective function converge.}
        \ENSURE $\mathbf{F}^{(i-1)}$,$\mathbf{w}^{(i-1)}$
    \end{algorithmic}
\end{algorithm}

\subsection{Auxiliary Variables Optimization}
With other parameters fixed, we can update the auxiliary variables $\boldsymbol{\xi}^c, \boldsymbol{\xi}^s$ for quadratic transform parameters by solving the following subproblem
\vspace{-0.2cm}
\begin{equation}
    (\text{SP.2}) \max_{\boldsymbol{\xi}^s,\boldsymbol{\xi}^c}\hat{\mathcal{G}}(\boldsymbol{\xi}^s,\boldsymbol{\xi}^c|\mathbf{F},\mathbf{x}, \mathbf{y}, \mathbf{w}, \boldsymbol{\mu})~\quad \\ .\label{eq:xisxic}
\end{equation}
\vspace{-0.1cm}
Given that $\hat{\mathcal{G}}(\boldsymbol{\xi}^s,\boldsymbol{\xi}^c|\mathbf{F},\mathbf{x}, \mathbf{y}, \mathbf{w}, \boldsymbol{\mu})$ is concave w.r.t. $\boldsymbol{\xi}^s$ and $\boldsymbol{\xi}^c$, we can obtain the closed-form solutions by setting the patio derivatives of $\boldsymbol{\xi}^s$ and $\boldsymbol{\xi}^c$ to zero, i.e., $\frac{\partial \hat{\mathcal{G}}(\boldsymbol{\xi}^s,\boldsymbol{\xi}^c|\mathbf{F},\mathbf{x}, \mathbf{y}, \mathbf{w}, \boldsymbol{\mu})}{\partial \xi^c} = 0 $ and $ \frac{\partial\hat{\mathcal{G}}(\boldsymbol{\xi}^s,\boldsymbol{\xi}^c|\mathbf{F},\mathbf{x}, \mathbf{y}, \mathbf{w}, \boldsymbol{\mu})}{\partial \xi^s} = 0$. The resulting closed-form solutions are thus obtained as follows
\begin{equation}
        \boldsymbol{\xi}^{s} = \frac{\sqrt{1+\mu_{K+1}}\left(\sqrt{\eta_s}\alpha_s \mathbf{w}^H \mathbf{b}_s(\mathbf{y})\mathbf{a}_s^H(\mathbf{x})\mathbf{F}\right)^*}{A},
        \label{ksis_update}
\end{equation}
\vspace{-0.8cm}
\begin{align}
    A=\sum_{c=1}^C&||\sqrt{\eta_c}\alpha_c \mathbf{w}^H  \mathbf{b}_c(\mathbf{y}) \mathbf{a}_c^H(\mathbf{x})\mathbf{F}||^2+||\mathbf{w}^H H_{\text{SI}}(\mathbf{x},\mathbf{y})\mathbf{F}||^2  \nonumber\\
    &+||\sqrt{\eta_s}\alpha_s \mathbf{w}^H \mathbf{b}_s(\mathbf{y})a_s^H(\mathbf{x})\mathbf{F}||^2+\|\mathbf{w}^H\|^2 \mathbf{\sigma}_s^2.
\end{align}
Similarly, the closed-form solution for $\xi^{c}_k$ can be derived as
\begin{equation}
    \xi^{c}_k = \frac{\sqrt{1+\mu_{k}}\mathbf{f}_k^H \mathbf{h}_k \left(\mathbf{x}\right)}{\sum_{j=1}^{K} \left|\mathbf{h}_k^H  \left(\mathbf{x}\right) \mathbf{f}_j \right|^2 + \sigma_k^2}.
    \label{ksic_update}
\end{equation}
Next, we address the subproblem of optimizing $\boldsymbol{\mu}$

\begin{equation}
    (\text{SP.3}) \max_{\boldsymbol{\mu}}\hat{\mathcal{G}}(\boldsymbol{\mu}|\mathbf{F},\mathbf{x}, \mathbf{y}, \mathbf{w}, \boldsymbol{\xi}^s ,\boldsymbol{\xi}^c)~\quad \\ .
    \vspace{-0.2cm}
\end{equation}
This can be solved by setting the derivative of $\boldsymbol{\mu}$ to zero, yielding
\begin{equation}
\mu_k = \frac{(B_k)^2 + B_k \sqrt{(B_k)^2 + 4}}{2},  k \in \{1, \dots, K+1\},
\vspace{-0.15cm}
\label{mu_dupdate}
\end{equation}
where $B_k = \text{Re}\left\{ \xi^{c}_k \mathbf{h}^{H}_k(\mathbf{x})\mathbf{f}_k(\mathbf{x}) \right\}, k = \{1, \dots, K\}
$ 
and $B_{K+1} = \text{Re} \left\{ \sqrt{\eta_s}\alpha_s \mathbf{w}^H \mathbf{b}_s(\mathbf{y}) \mathbf{a}^H_s(\mathbf{x}) \mathbf{F} \boldsymbol{\xi}^s \right\}$.
Consequently, the algorithm for obtaining locally optimal solutions of the transceivers' beamforming matrices and the introduced auxiliary variables is summarized in \textbf{Algorithm \ref{algo1}}.

\vspace{-0.2cm}
\subsection{Antenna Position Optimization}
The antenna positions for both $\mathbf{x}$ and $\mathbf{y}$ can be updated by solving the respective subproblems.
\begin{equation}
    (\text{SP.4}) \max_{\mathbf{x}}\hat{\mathcal{G}}(\mathbf{x}|\mathbf{F},\mathbf{y}, \mathbf{w}, \boldsymbol{\mu},\boldsymbol{\xi}^c,\boldsymbol{\xi}^s)~\quad \\ \text{s.t.} \ \text{(\ref{eq:xy1}),\ (\ref{eq:xy2})}.
\end{equation}
\begin{equation}
    (\text{SP.5}) \max_{\mathbf{y}}\hat{\mathcal{G}}(\mathbf{y}|\mathbf{F},\mathbf{x}, \mathbf{w}, \boldsymbol{\mu},\boldsymbol{\xi}^c,\boldsymbol{\xi}^s)~\quad \\ \text{s.t.} \ \text{(\ref{eq:xy1}),\ (\ref{eq:xy2})}.
\end{equation}
Given the problem's non-convexity, obtaining optimal solutions is challenging. Thus, we propose a two-stage approach combining coarse and fine granularity methods to update antenna position.

A coarse search is first performed over the grid location sets $\mathcal{S}_X$ and $ \mathcal{S}_Y$ to determine suitable initialization positions. These sets consist of points starting from $x = X_{\min}$ and $y = Y_{\min}$ with a sampling interval $\lambda$ within the movable range. During this coarse-grained search, $N_T$ points are selected from all possible subsets of $\mathcal{S}_X$, while $N_R$ points are selected from $\mathcal{S}_Y$. The objective function is evaluated after running \textbf{Algorithm \ref{algo1}} for five iterations to reduce computational complexity. After assessing all possible combinations, the set that maximizes the objective function is selected as the initial values for $\mathbf{x}$ and $\mathbf{y}$. Subsequently, fine-grained adjustments on the best initial points are conducted using the gradient projection method\cite{securemovable,lyu2024flexible}. The antenna positions $x_n$, $n\in\{1,...,N_T\}$, $y_m$, $m\in\{1,...,N_R\}$ can be alternatively updated as 
\begin{equation}
    x_n^{(i+1)}=x_n^{(i)}+\delta^t\nabla_{x_n}\hat{\mathcal{G}}(\mathbf{x}|\mathbf{F},  \mathbf{y}, \mathbf{w}, \boldsymbol{\mu},\boldsymbol{\xi}^c,\boldsymbol{\xi}^s),
    \label{x_update}
\end{equation}
\begin{equation}
    y_m^{(i+1)}=y_m^{(i)}+\delta^t\nabla_{y_m}\hat{\mathcal{G}}(\mathbf{y}|\mathbf{F},  \mathbf{x}, \mathbf{w}, \boldsymbol{\mu},\boldsymbol{\xi}^c,\boldsymbol{\xi}^s),
    \label{y_update}
\end{equation}
where $i$ denotes the iteration number of the inter-loop for antenna position optimization and $\delta^t$ denotes the step size of the gradient ascent method. Next, we project to meet (\ref{eq:xy1}), (\ref{eq:xy2}). The update process for the receive antenna positions is similar to that of transmitter. For simplicity, the explanation will focus solely on the transmit antenna positions. The antenna indices are rearranged as  
$X_{\min} \leq \hat{x}_1 \leq \hat{x}_2 \leq \cdots \leq \hat{x}_{N_T} \leq X_{\max}$. The final step involves projecting onto the feasible region, which entails sorting the updated values of $\mathbf{x}$ after the last round of gradient ascent in ascending order, reassigning antenna indices accordingly, and then adjusting the antenna spacing. The locally optimal antenna positions are then determined following this projection
\begin{equation}
\scalebox{0.83}{$
\begin{cases}
\hat{x}_1^{t+1} = \max \left( X_{\min}, \min \left( X_{\max} - (N - 1)D_0, \hat{x}_1^{t+1} \right) \right), \\
\hat{x}_2^{t+1} = \max \left( \hat{x}_1^{t+1} + D_0, \min \left( X_{\max} - (N - 2)D_0, \hat{x}_2^{t+1} \right) \right), \\
\hdots \\
\hat{x}_N^{t+1} = \max \left(\hat{x}_{N-1}^{t+1} + D_0, \min \left( X_{\max},\hat{x}_N^{t+1} \right) \right).
\label{eq:antenna projection}
\end{cases}
$}
\vspace{-0.2cm}
\end{equation}
\begin{algorithm}[t]
    \renewcommand{\algorithmicrequire}{\textbf{Initialization:}}
	\renewcommand{\algorithmicensure}{\textbf{Output:}}
    \caption{Proposed CFGS algorithm for updating the antenna positions at Tx and Rx.}
     \label{algo2}
    \begin{algorithmic}[1]
        \REQUIRE Generate all the possible position alignments of transmit antenna as $\{ \bm{\zeta}_{x1}, \bm{\zeta}_{x2}, \cdots, \bm{\zeta}_{xq_{x}} \}$ from $\mathcal{S}_{X}$. And all the possible position alignments of receive antenna as $\{ \bm{\zeta}_{y1}, \bm{\zeta}_{y2}, \cdots, \bm{\zeta}_{yq_{y}} \}$ from $\mathcal{S}_{Y}$, set iteration index $l=1$.
        \FOR{$i=1,2,\cdots, q_{x}$}
            \FOR{$j=1,2,\cdots, q_{y}$}
            \STATE Let $\mathbf{x}=\bm{\zeta}_{xi}$ , $\mathbf{y}=\bm{\zeta}_{yj}$. 
            \STATE Converge $\mathbf{F}^{(ij)}$ and $\mathbf{w}^{(ij)}$ with \textbf{Algorithm \ref{algo1}}. 
            \STATE Compute $R_{ij}=\mathcal{G}(\mathbf{F}^{(ij)}, \mathbf{x}, \mathbf{y}, \mathbf{w}^{(ij)})$.
            \ENDFOR
        \ENDFOR
        \STATE Let $\mathbf{x}^{(0)}={\bm{\zeta}}_{xk}, \mathbf{y}^{(0)}={\bm{\zeta}}_{yt}, k,t=\arg\max R_{kt}$ .
        \REPEAT
            \STATE Converge $\mathbf{F}^{(l)}$ and $\mathbf{w}^{(l)}$ with \textbf{Algorithm \ref{algo1}}.
            \STATE Converge $\mathbf{x}^{(l)}$ and $\mathbf{y}^{(l)}$ as (\ref{x_update}) and (\ref{y_update}). 
            \STATE Adjust $\mathbf{x}^{(l)}$ and $\mathbf{y}^{(l)}$ as (\ref{eq:antenna projection}).
            \STATE Update iteration index $l=l+1$.
        \UNTIL{the value of objective function converge.}
        \ENSURE $\mathbf{x}^{(l-1)}$ and $\mathbf{y}^{(l-1)}$ 
    \end{algorithmic}
\end{algorithm}

Based on the subproblems discussed, the overall algorithm to solve (\(\text{P1}\)) is summarized in \textbf{Algorithm \ref{algo2}}. Since \(\hat{\mathcal{G}}(\mathbf{F}, \mathbf{x}, \mathbf{y}, \mathbf{w}, \boldsymbol{\mu}, \boldsymbol{\xi}^c, \boldsymbol{\xi}^s)\) is non-decreasing with each iteration and has an upper bound, \textbf{Algorithm \ref{algo2}} is guaranteed to converge. The computational complexity of \textbf{Algorithm \ref{algo1}} is approximately \(\mathcal{O} \big(T_1\big(KN_T^3 + KN_RN_T + KN_R^2 + C(N_TK + N_R)\big)\big)\), where \(T_1\) is the number of iterations for updating \(\mathbf{F}\) and \(\mathbf{w}\). Letting \(I_1\) and \(I_2\) denote the number of coarse-grained and fine-grained search iterations, respectively, and \(T_2\) denotes the number of iterations for updating \(\mathbf{x}\) and \(\mathbf{y}\), the overall complexity is \(\mathcal{O} \big((I_1 + I_2)T_1\big(KN_T^3 + KN_RN_T + KN_R^2 + C(N_TK + N_R)\big) + I_2T_2(C(N_R + KN_T) + K^2N_T + KN_RN_T)\big)\).

\section{Simulation Results}
Two schemes are compared: fixed position antenna (\textbf{FPA}) and gradient ascent with movable antenna (\textbf{GA-MA}). In the FPA method, the antenna spacing of the transmit and receive antennas are $\lambda/2$ and beamforming matrices are optimized as \textbf{Algorithm \ref{algo1}}. In the GA-MA method, the transmit and receive antennas are initially randomly located in the movable range and directly optimized with the gradient ascent method as shown from \textbf{Step 9} to \textbf{Step 14} in \textbf{Algorithm \ref{algo2}}.

We consider paths number $L_p = 12$, $K = 4$ users, and $C = 3$ clutters. Users and clutters are randomly positioned around the BS within the angle range $[0,\pi]$, with the sensing target direction fixed at $\pi/4$. User-BS distances are randomly distributed within $[50\,\text{m}, 80\,\text{m}]$, while the target-BS distance ranges within $[10\,\text{m}, 20\,\text{m}]$. The complex RCS coefficients and the channel gain follow the standard complex Gaussian distribution, i.e., $\alpha_{s},\alpha_{c},\rho_{k,l} \sim \mathcal{C}\mathcal{N}(0,1)$. Given a carrier frequency of 30 GHz, the wavelength is $\lambda=0.01$ m. The free-space path losses for users, target, and clutters channels are set as $\eta=\big[\frac{\sqrt{G_l\lambda}}{4\pi d}\big]^2$ and $G_l=1$ as the BS is equipped with omnidirectional antennas. The received noise power is $-60$ dBm. The feasible lower bounds for MAs $X_{\min}$ and $Y_{\min}$ are set to 0 and $D_0=\frac{\lambda}{2}$, respectively.
\begin{figure}[t]
    \centering
    \includegraphics[width=0.72\linewidth]{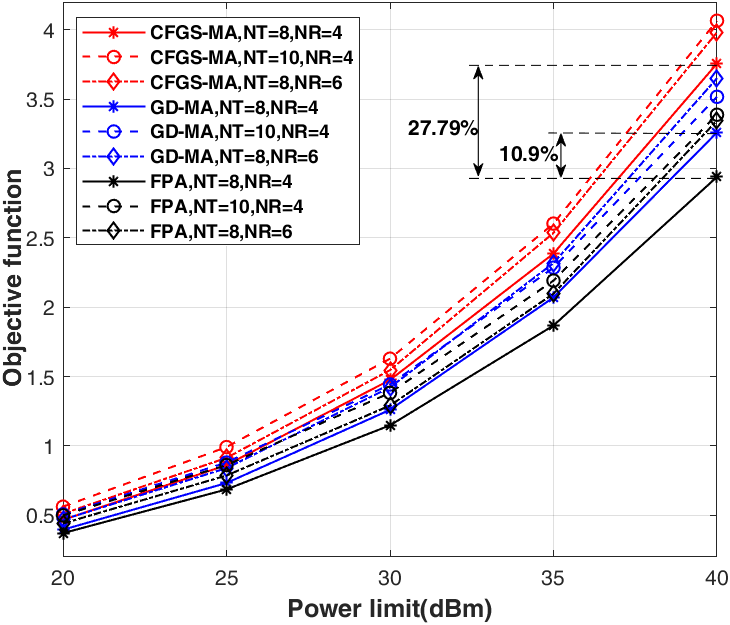}
    \caption{ISAC performance with different transmit power, $X_{\max}=12\lambda$, $Y_{\max}=8\lambda$, $\varpi_s=0.5$ and $\varpi_c=0.5$, $r_{0}=20\lambda$.}
   \vspace{-0.45cm}
    \label{fig:power_comparison}
\end{figure}
Fig. \ref{fig:power_comparison} illustrates the ISAC performance across varying transmit power levels and different antenna configurations. The transmit power ranges from 20 dBm to 40 dBm, with three sets of antenna number configurations used in the simulation: $\{8,4\}$, $\{10,4\}$, and $\{8,6\}$ for transmit and receive antennas, respectively.
The result shows that the objective function grows with increasing transmit power and the number of antennas. When transmit power $P_0$ = 40 dBm, 8 transmit antennas and 4 receive antennas are used. The objective function increases by 10.9\% with GA-MA compared to FPA, and a 27.79\% increment when CFGS-MA is used compared to FPA.
Furthermore, for all three sets of antenna number in the simulation, CFGS-MA shows a better performance than GA-MA and FPA.    

\begin{figure}[t]
    \centering
    \includegraphics[width=0.77\linewidth]{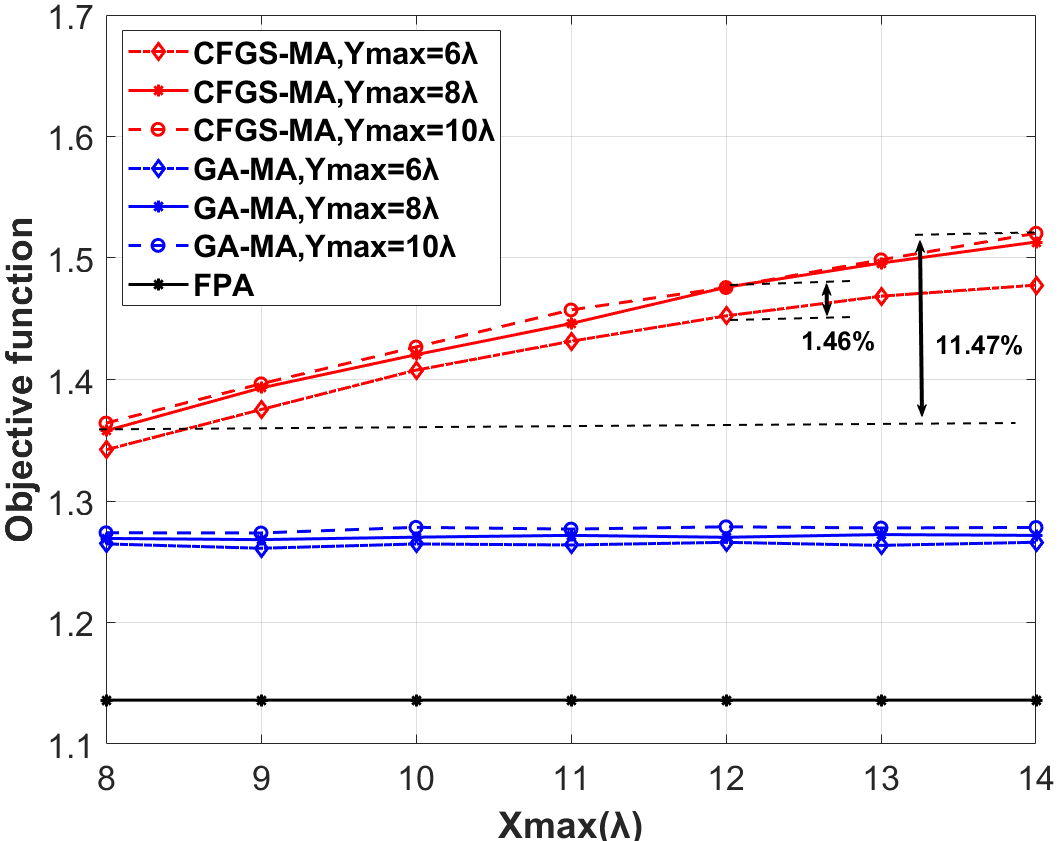}
    \caption{ISAC performance with different antenna movable range, $P_0=30$ dBm, $N_T=8$, $N_R=4$, $\varpi_s=0.5$ and $\varpi_c=0.5$, $r_{0}=20\lambda$.}
    \vspace{-0.3cm}
    \label{fig:range_comparison}
\end{figure}

Fig. \ref{fig:range_comparison} shows the ISAC performance with different movable range of transmit and receive antennas. The movable range of the transmit antennas varies from 8 $\lambda$ to 14 $\lambda$,  while the receive antennas are tested with three different movable ranges: 6 $\lambda$, 8 $\lambda$, and 10 $\lambda$. The results show that with GA-MA, performance improvement is small, as it only achieves a local optimum near the initial position, limiting the benefit of a larger movable range. In contrast, the proposed CFGS-MA significantly enhances performance, with an 11.47\% increase in the objective function as the transmit movable range expands from 8 $\lambda$ to 14 $\lambda$ with the receive movable range is 8 $\lambda$. Additionally, there is a 1.46\% increase as the movable range of receive antennas varies from 6 $\lambda$ to 10 $\lambda$ while maintaining transmit antennas' movable range at 12 $\lambda$. These results demonstrate that CFGS-MA effective utilization of the available range.

\begin{figure}[t]
    \centering
    \includegraphics[width=0.77\linewidth]{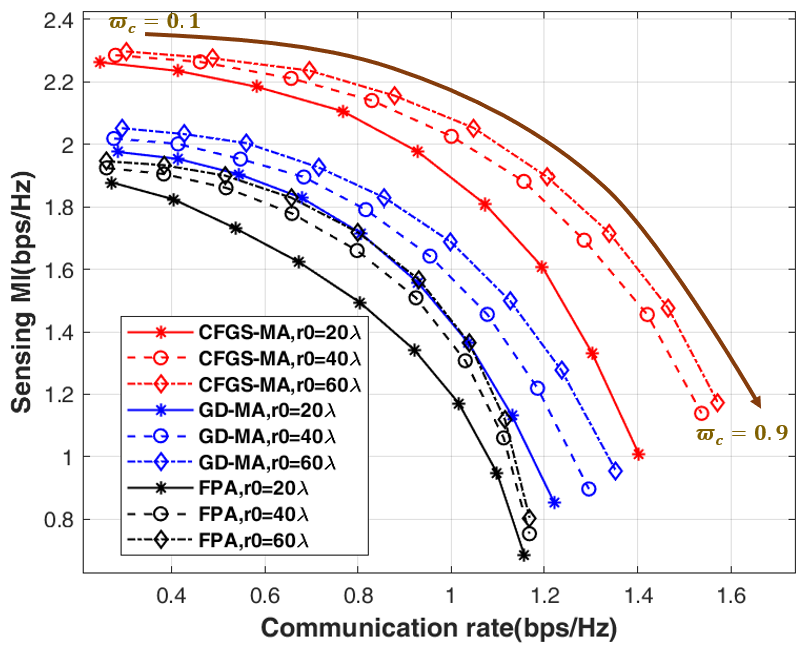}
    \caption{Trade-off between communication and sensing with different SI, $P_0=30$ dBm, $N_T=8$, $N_R=4$, $X_{\max}=12\lambda$ and $Y_{\max}=8\lambda$.}
    \vspace{-0.3cm}
    \label{fig:rate_comparison}
\end{figure}

Fig. \ref{fig:rate_comparison} shows the trade-off between communication and sensing by varying $\varpi_c$ from 0.1 to 0.9. The results indicate that CFGS-MA consistently outperforms GA-MA and FPA across different trade-off scenarios. Additionally, the impact of different $d_{Tx,Rx}$ values, which influence self-interference, is compared. Among the three methods, CFGS-MA demonstrates superior performance under varying levels of interference.

\vspace{-0.1cm}
\section{Conclusion}

This paper focuses on maximizing the communication rate and sensing MI in an FD monostatic MA-ISAC system. We model the SI channel as a function of the antenna position vectors under near-field channel condition, enabling effective self-interference suppression. To address the non-convexity of the formulated problem, we apply the FP and AO methods to transform the problem into multiple subproblems. In particular, for the optimization of antenna positions, we proposed a two-stage CFGS algorithm. Numerical results demonstrated that our proposed CFGS offers significant advantages over GA-MA and FPA configurations under various conditions.